\documentclass[journal]{IEEEtran}
\ifCLASSINFOpdf
  \usepackage[pdftex]{graphicx}
  \DeclareGraphicsExtensions{.pdf,.jpeg,.png}
\else
  \usepackage[dvips]{graphicx}
  \DeclareGraphicsExtensions{.eps}
\fi
\usepackage{cite}
\usepackage{amsmath}
\usepackage{amssymb}
\usepackage{breqn}
\usepackage{multirow}
\usepackage[official]{eurosym}
\usepackage{mleftright}
\usepackage{enumitem}
\usepackage{nomencl}
\usepackage{color}
\makenomenclature

\mleftright
\usepackage{array}

\usepackage{color}

\begin{document}
\title{A New Cooperative Framework for a Fair and Cost-Optimal Allocation of Resources within a Low Voltage Electricity Community}
\author{Martin~Hupez,
        Jean-François~Toubeau,~\IEEEmembership{Member,~IEEE,}
        Zacharie~De~Grève,~\IEEEmembership{Member,~IEEE,}
        and~François~Vallée,~\IEEEmembership{Member,~IEEE}
\thanks{The authors were with the Power Systems \& Markets Research Group, Electrical Power Engineering Unit, University of Mons, B-7000 Mons, Belgium. E-mail: martin.hupez@umons.ac.be}
\thanks{Accepted on 11/19/2020 for publication in IEEE Transactions on Smart Grid. \copyright 2020 IEEE. Personal use of this material is permitted. Permission from IEEE must be obtained for all other uses, in any current or future media, including reprinting/republishing this material for advertising or promotional purposes, creating new collective works, for resale or redistribution to servers or lists, or reuse of any copyrighted component of this work in other works.}}
\maketitle

\begin{abstract}
This paper presents an original collaborative framework for power exchanges inside a low voltage community. The community seeks to minimize its total costs by scheduling on a daily basis the resources of its members. In this respect, their flexibility such as excess storage capacity, unused local generation or shiftable load are exploited. Total costs include not only the energy commodity, but also grid fees associated to the community operation, through the integration of power flow constraints. In order to share the community costs in a fair manner, two different cost distributions are proposed. The first one adopts a distribution key based on the Shapley value, while the other relies on a natural consensus defined by a Nash equilibrium. Outcomes show that both collaboration schemes lead to important savings for all individual members. In particular, it is observed that the Shapley-based solution gives more value to mobilized flexible resources, whereas the Nash equilibrium rewards the potential flexibility consent of end-users. 
\end{abstract}


\nomenclature[V]{$x_{n,a}^t$}{Power of appliance $a$ scheduled at $t$ by user $n$,}
\nomenclature[V]{$s_{n,\mathrm{ind}}^t$}{Power of locally stored energy at $t$ by user $n$,}
\nomenclature[V]{$e_{n,\mathrm{ind}}^t$}{Energy of user $n$ stored locally at $t$,}
\nomenclature[V]{$s_{n,\mathrm{host}}^t$}{Power of hosted energy at $t$ by user $n$,}
\nomenclature[V]{$e_{n,\mathrm{host}}^t$}{Energy hosted at $t$ by user $n$,}
\nomenclature[V]{$e_{n,\mathrm{tot}}^t$}{Total energy stored in user $n$'s battery,}
\nomenclature[V]{$s_{n,\mathrm{mut}}^t$}{Power of remotely stored energy at $t$ by user $n$,}
\nomenclature[V]{$e_{n,\mathrm{mut}}^t$}{Energy stored remotely at $t$ by user $n$,}
\nomenclature[V]{$d_n^t$}{Non-flexible load power of user $n$ forecasted at $t$,}
\nomenclature[V]{$v_n^t$}{Free excess power allocated to user $n$ at $t$,}
\nomenclature[V]{$l_n^t$}{Virtual net load of user $n$ at $t$,}
\nomenclature[V]{$l_n^{t+}$}{Virtual net load absolute positive value of $n$ at $t$,}
\nomenclature[V]{$l_n^{t-}$}{Virtual net load absolute negative value of $n$ at $t$,}
\nomenclature[V]{$\pi_{n,a}^t$}{Permissibility binary of appliance $a$ consented by $n$,}
\nomenclature[V]{$\delta_{j,k}$}{Electrical distance between nodes $j$ and $k$,}
\nomenclature[V]{$p_n$}{Active power injected at node $n$,}
\nomenclature[V]{$q_n$}{Reactive power injected at node $n$,}
\nomenclature[V]{$p_{n,j}$}{Active branch power flowing from $n$ to $j$,}
\nomenclature[V]{$q_{n,j}$}{Reactive branch power flowing from $n$ to $j$,}
\nomenclature[V]{$\phi_{n,j}$}{Squared branch current magnitude from $n$ to $j$,}
\nomenclature[V]{$w_n$}{Squared voltage at node $n$,}
\nomenclature[P]{$R_{n,j}$}{Resistance of the branch connecting $n$ and $j$,}
\nomenclature[P]{$X_{n,j}$}{Reactance of the branch connecting $n$ and $j$,}
\nomenclature[P]{$\gamma_{\mathrm{supp},i}^t$}{Commodity price of supplier $i$ at $t$,}
\nomenclature[P]{$\gamma_{\mathrm{up}}$}{Upstream grid fees,}
\nomenclature[P]{$\gamma_{\mathrm{loss}}$}{Price of losses within the local grid,}
\nomenclature[P]{$\gamma_{\mathrm{flow}}$}{Price of line power flows within the local grid,}
\nomenclature[S]{$\mathcal{N}$}{Set of community members,}
\nomenclature[S]{$\mathcal{N}_i$}{Set of supplier $i$'s members,}
\nomenclature[S]{$\mathcal{S}$}{Set of suppliers,}
\nomenclature[S]{$\mathcal{Q}$}{Coalition/subset of community members,}
\nomenclature[S]{$\mathcal{T}$}{Set of optimization intervals,}
\nomenclature[S]{$\mathcal{A}_n$}{Member $n$'s set of appliances,}
\nomenclature[P]{$E_{n,a}$}{Total predefined energy for appliance $a$ of member $n$,}
\nomenclature[P]{$M_{n,a}$}{Maximum power level for appliance $a$ of $n$,}
\nomenclature[P]{$M_{n}^{\mathrm{dis}}$}{Maximum power level for discharging $n$'s battery,}
\nomenclature[P]{$M_{n}^{\mathrm{ch}}$}{Maximum power level for charging $n$'s battery,}
\nomenclature[P]{$K_{n}$}{Maximum energy capacity of $n$'s battery,}
\nomenclature[P]{$\Delta t$}{Duration of optimization interval,}
\nomenclature[F]{$C_{\mathrm{supp},i}^t(.)$}{Commodity costs function of supplier $i$,}
\nomenclature[F]{$C_{\mathrm{up}}^t(.)$}{Upstream grid costs function,}
\nomenclature[F]{$C_{\mathrm{loc}}^t(.)$}{Local grid costs function,}
\nomenclature[F]{$\phi_n(.)$}{Shapley value function,}
\nomenclature[V]{$p_0^t$}{Power at the community interface node at $t$,}
\nomenclature[S]{$\Theta_n$}{Set of member $n$'s decisions variables,}
\nomenclature[S]{$\Theta$}{Set of all decisions variables,}
\nomenclature[S]{$\mathcal{Y}_n$}{Set of adjacent nodes connected to $n$,}
\nomenclature[S]{$\mathcal{B}$}{Set of all branches of the low voltage entity,}
\nomenclature[V]{$C_{\mathrm{supp},i}^{\{n\}*}$}{Optimum of $n$'s commodity costs,}
\nomenclature[V]{$C_{\mathrm{supp}}^{\{\mathcal{N}\}*}$}{Optimum community commodity costs,}
\nomenclature[V]{$C_{\mathrm{supp}}^{\{\mathcal{N}\}}$}{Actual community commodity costs,}
\nomenclature[V]{$C_{\mathrm{grid}}^{\{\mathcal{N}\}}$}{Actual community grid costs,}

\printnomenclature
\section{Introduction}

\IEEEPARstart{U}{nlocking} the flexibility distributed throughout the different levels of the electricity system appears to be a decisive step for a successful energy transition. The development of demand-side management programs, the integration of Distributed Energy Resources (DERs) and the deployment of smart grid technologies give new possibilities for the system, provided that the right incentives are put in place \cite{Klonari2016, Toubeau2015}. 

However, the large-scale adoption of such programs and technologies suffers the lack of a clear and coherent economic framework that will engage prosumers in a joint endeavour. The concept of consumer-centric systems aims at placing the end-users inside the economic cycle by enhancing their experience. Several visions concur for its application in power systems, each requiring different structures for the power exchanges \cite{Parag2016,Pinson2017}. In its most centralized form, it aims at extending the accessibility of the existing pool-based markets to all the prosumers and integrate them with the upstream actors \cite{Elia2018}. The efforts focus mainly on giving access to market-based signals to downstream participants, allowing price-responsive actors to optimize their bill and monetize their flexibility. This system, hence, requires enormous coordination efforts, which may necessitate significant bidirectional communication and may have privacy implications. Conversely, the most decentralized approach considers a peer-to-peer (P2P) market solution, in which consumers and producers can directly sell and buy electricity as well as related services without centralized supervision \cite{Sousa2019, Morstyn2019a,Kang2017}. In this way, they can choose how they source their electric energy, hence turning it into a product with a high potential of differentiation \cite{Morstyn2018b,Sorin2019}. By coordinating P2P transactions, Morstyn et al. \cite{Morstyn2018a} argue that it is also possible to address higher-level needs such as balancing services. However, this decentralized design relies on the complex task of properly incentivizing coordination (towards improved network efficiency) between end-users, each with different economic resources and motivations, potentially characterized by a non-rational behavior. The problem is further complicated by the inherent disparities in terms of network infrastructure. For instance, the P2P transactions fees proposed in \cite{Morstyn2020} and based on Distribution locational marginal prices discriminate prosumers based on their location on the grid. An intermediate and more practical solution consists thus in forming communities. The latter are structured around a collective of prosumers that share common interests and goals \cite{Sousa2019}, which result from a mix of subjective considerations (e.g. maximum autonomy) and/or more objective reasons (e.g. geographical proximity). Most of the current literature considers the implementation of local markets supervised by a community manager \cite{Moret2018, Cai2017}, who is responsible of both the auctions inside the local market and the power exchanges outside the community.

Overall, acting at community level offers a good compromise between centralized and fully distributed approaches. However, introducing markets in such small entities is subject to some caveats since the main assumptions of perfect competition are hardly met. For instance, the limited size of community-based markets tends to inherently give market power to some individuals.
Besides, modeling assets utility at such levels assumes that the demand is elastic, whereas it is hard to conceive that residential prosumers or small businesses would significantly modify their energy needs on a daily basis. 

In order to avoid issues related to market-based transactions, an alternative approach to enable consumer-centricity is to build communities in which cooperation prevails over competition. Such a solution consists in sharing the aggregated flexibility in a fair and optimal manner based on objective costs, needs and constraints, and can be perceived as a concept enhancing costs-awareness and virtuous interactions. 
Various contributions address these considerations in scenarios involving groups of interacting members \cite{Mohsenian-Rad2010a,Hupez2018,Atzeni2012,Baharlouei2013,Stevanoni2019,Lee2014,Han2019}. Among them, game theoretical approaches are commonly used when strategic behaviors are possible. On the one hand, non-cooperative games are formulated when interactions are not subject to binding agreements \cite{Mohsenian-Rad2010a,Hupez2018,Atzeni2012,Baharlouei2013, Stevanoni2019}. By designing the game appropriately, it is possible to make cooperation self-enforcing through its Nash equilibrium. In \cite{Mohsenian-Rad2010a}, an aggregative game solving the day-ahead energy consumption scheduling for a set of users owning time flexible appliances is formulated. The problem is solved using an autonomous and distributed best response algorithm, and it is shown that the outcome is equivalent to the one of the global optimization. The approach is complemented in \cite{Hupez2018} by considering a liberalized framework with multiple suppliers and a Distribution System Operator (DSO) who manages grid-related aspects. These formulations, however, lack the possibility of incentivizing participants to consent higher flexibility. The introduction of DERs combined with the use of an hourly distribution of the costs in \cite{Atzeni2012} address this issue. The Nash equilibrium is computed using a proximal decomposition algorithm, but higher flexibility is obtained at the cost of a non-optimal global solution \cite{Baharlouei2013}. On the other hand, cooperative game theory focuses on incentivizing the formation of coalitions \cite{Lee2014,Han2019}. In such schemes, cooperation is based on a predefined consensus enforced by a supervizing authority. In \cite{Lee2014}, end-users form coalitions for trading a portion of their electricity. Their participation is optimized by relying on the Shapley value for adequately distributing energy savings. Han et al. \cite{Han2019} propose a similar approach, which is complemented with an energy management optimization. 
Such a predefined consensus allows to explicitly define how costs are distributed among end-users, thus ensuring fairness within the community. However, two shortcomings still need to be addressed in the current literature.
Firstly, current works focus on commodity costs, and neglect network-related fees, while the latter contribution represents the highest expenditures within the electricity bill. Secondly, they fail to consider that the local generation and flexibility can be valued at another price than the utility costs within the community (e.g. the truth marginal cost of photovoltaic generation is near zero), which may limit the mutualization potential of the available resources.

In this context, this paper proposes an original day-ahead cooperative framework aiming to minimize both commodity and network costs in a low voltage entity, without resorting to a complex market structure. A fair distribution of the cost savings is targeted to leverage the involvement of all its members and enhance the sharing of DERs. Practically, we compare two different distribution keys, namely a natural consensus arising from the Nash equilibrium, and a solution provided by the Shapley value. 
The proposed framework fits naturally to a distributed resolution such as the Alternating Direction Method of Multipliers \cite{Boyd2010}, hence ensuring that no third-party is necessary for the reliable operation of the proposed design. The transparent computation and supervision (required to guarantee the safety and effectiveness of the proposed design) could be given by a blockchain and smart contracts \cite{Munsing2017}, such that the proposed concept naturally conforms within the current liberalized framework. Overall, the main contributions of this paper are the following:

\begin{itemize}
\item We establish a community framework where DERs and flexible resources are naturally mutualized by recording virtual power flows. The excess of generation and storage pertaining to individual end-users is thus made available to the rest of the community. This solution complies with the current liberalized context and raises engagement opportunities at the local level, thereby improving the cost-effectiveness of traditional strategies where each individual only considers its own electricity.
\item We formulate a day-ahead power exchange planning that minimizes the global community costs. The procedure accounts for grid fees and losses through the Second Order Cone Programming (SOCP) relaxation of power flow equations (DistFlow). This also allows for the enforcement of a scheduling solution that complies with the network's technical requirements. Furthermore, by considering the non-linear and aggregative nature of the grid costs, we acknowledge that the grid is a shared asset whose operation can be optimized through cooperation within the community.
\item We apply two different cost distributions to ensure the fairness in the final electricity bill of end-users. The goal is to allocate among individuals the costs savings from the mutualized flexibility in a transparent and rational manner. In particular, outcomes from the case study show that the consensus given by the Nash equilibrium tends to value the potential flexibility consented by end-users, while the Shapley value better rewards the actual mobilized flexibility.
\end{itemize}
This combination of a cost structure reflecting more accurately the use of resources with a framework enhancing higher engagement potential and a fair cost allocation to ensure the achievement of the scheme gives a holistic and innovative solution to help fulfill the energy transition objectives on the demand-side. The rest of the paper is structured as follows. Section \ref{sec:community} lays out the proposed community framework. In particular, the load, network and cost models are successively introduced. The mechanisms of the proposed design, i.e. the optimization problem and the cost distributions, are presented in Section \ref{sec:PES}. Finally, Section \ref{sec:case_studies} presents a case study and highlights interesting observations, before the conclusions and perspectives are drawn in Section \ref{sub:Conclusion}.

\section{Community framework}
\label{sec:community}
In order to adopt a scenario in which network considerations are fully integrated, it is assumed that community members are connected to the same low voltage (LV) entity (such as a distribution feeder or a collection of them). The proposed design can be seamlessly integrated into the current liberalized framework: the community network is managed by a single DSO, while each member can choose among different suppliers for its net consumption $l_{n}^{t}$ (accounting for the mutualization of resources with other end-users). Besides, note that curly inequality operators (e.g. $\succeq$) are used for component-wise inequalities and the delta-equal-to symbols ($\triangleq$) define variables or sets. 

\begin{figure}[!b]
\centering
\includegraphics[scale=0.36]{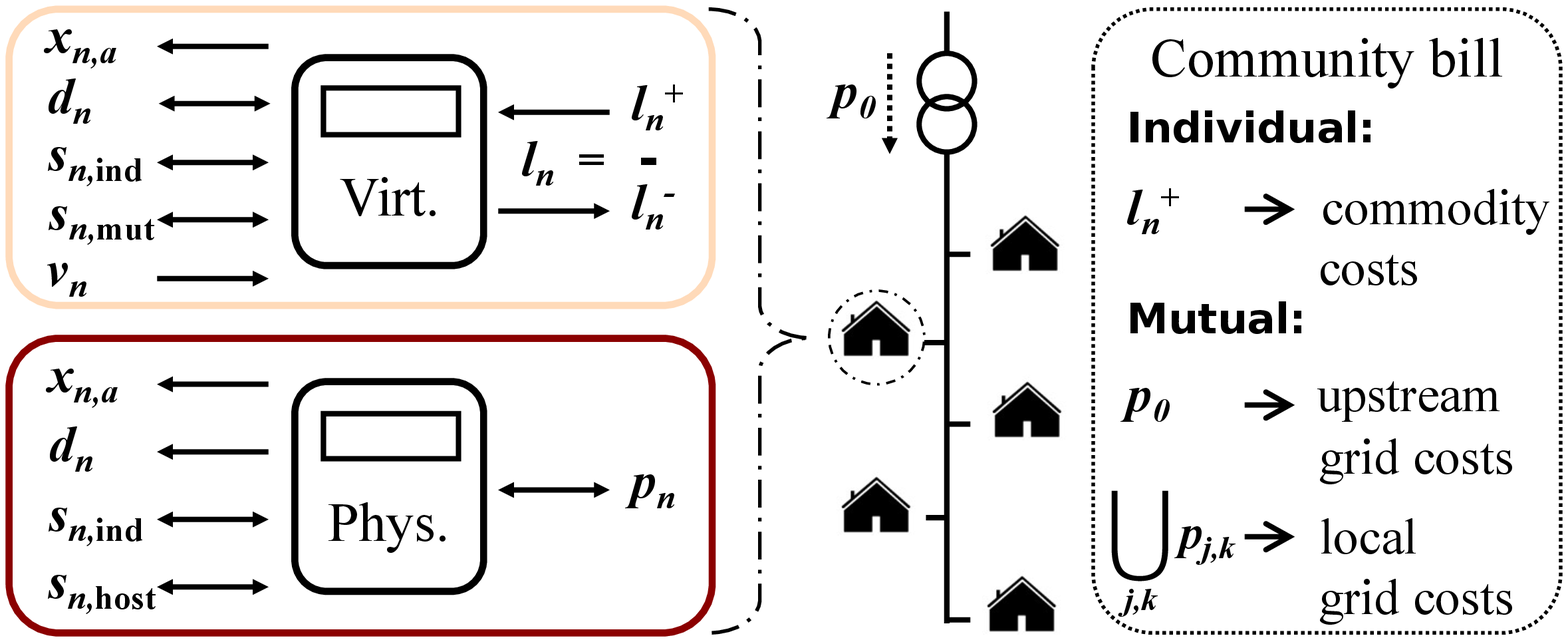}
\caption{Community bill formation based on virtual and physical flows.}
\label{fig:community_bill}
\end{figure}

Let $n \in \mathcal{N}\triangleq\{1,2,\ldots,N\}$ denote the set of community members, and $t \in \mathcal{T} \triangleq\{1,\ldots,T\}$ the set of optimization intervals considered for one day of scheduling, each of duration $\Delta t$. To enable more efficient power exchanges through mutualization of all individual resources, the electricity flows are differentiated between virtual and physical flows (Fig.~\ref{fig:community_bill}). Virtual flows $l_{n}^{t}$ record the energy transactions of the prosumers for billing purposes, and do not necessarily reflect the actual physical flows within the network. Obviously, the sum of all virtual and physical flows are equal.

\subsection{Prosumer load model}
For each community member $n\in\mathcal{N}$, we can distinguish different load components. We firstly define the flexible consumption, i.e. the portion of the load for which prosumers consent flexibility in their operation (e.g. washing machine, heat pump, electric vehicle, etc.). This set of appliances, denoted by $\mathcal{A}_n$, can be operated at the most opportune time under intertemporal constraints (some processes have fixed load profiles and/or cannot be stopped during operation). Each appliance $a\in\mathcal{A}_n$ is described by a power scheduling vector: 
\begin{equation}
\mathbf{x}_{n,a} \triangleq \left[ x_{n,a}^1 ,\ldots, x_{n,a}^T \right] \succeq 0\text{,}
\end{equation} 
which is formed by the collection of the corresponding values at time intervals $t\in \mathcal{T}$. 

Secondly, the non-flexible load includes critical appliances (e.g. lighting, kitchen equipment, multimedia devices, etc.) as well as the non-dispatchable local generation, that is, energy produced by photovoltaic (PV) panels and small wind turbines, counted negatively. Its forecasted power vector \cite{Toubeau2019} is defined by:
\begin{equation}
\mathbf{d}_{n} \triangleq \left[ d_{n}^1 ,\ldots, d_{n}^T \right]\text{.}
\end{equation}

Thirdly, prosumers can have a personal Energy Storage System (ESS) such as a battery. The power scheduling vector $\mathbf{s}_{n,\mathrm{ind}}$ for their own utilization is:
\begin{equation}
\mathbf{s}_{n,\mathrm{ind}} \triangleq \left[ s_{n,\mathrm{ind}}^1 ,\ldots, s_{n,\mathrm{ind}}^T \right]\text{,}
\end{equation}
where a positive value stands for the charging mode. Then, each member makes its excess storage space available to other members in need. The power hosted $\mathbf{s}_{n,\mathrm{host}}$ is defined by:
\begin{equation}
\mathbf{s}_{n,\mathrm{host}} \triangleq \left[ s_{n,\mathrm{host}}^1 ,\ldots, s_{n,\mathrm{host}}^T \right]\text{.}
\end{equation}

Therefore, the end-users can possibly benefit from the excess of storage space pooled by the other members of the community. The corresponding power scheduling vector $\mathbf{s}_{n,\mathrm{mut}}$ of shared ESS space allocated to member $n$ is:
\begin{equation}
\mathbf{s}_{n,\mathrm{mut}} \triangleq \left[ s_{n,\mathrm{mut}}^1 ,\ldots, s_{n,\mathrm{mut}}^T \right]\text{.}
\end{equation}

Similarly, the excess of power generated (i.e. outflows of end-users) within the community is made available to other members. The portion allocated to member $n$ is given by:
\begin{equation}
\mathbf{v}_{n} \triangleq \left[ v_{n}^1 ,\ldots, v_{n}^T \right] \succeq 0\text{.}
\end{equation}

Hence, the individual virtual net load (which is referred to as net load in the remainder of the paper) metered by the DSO at time step $t$ (Fig. \ref{fig:community_bill}) is defined as:
\begin{equation}
l_n^t = d_n^t + \sum_{a\in\mathcal{A}_{n}} x_{n,a}^t + s_{n,\mathrm{ind}}^t - v_n^t + s_{n,\mathrm{mut}}^t\text{.} \label{eq:netcons}
\end{equation}

Additionally, the load scheduling variables are subject to individual requirements and physical constraints. We thereby characterize the scheduling flexibility for each appliance $a$ of member $n$ by a permissibility vector: 
\begin{equation}
\boldsymbol{\pi}_{n,a} \triangleq \left[\pi_{n,a}^1,\ldots,\pi_{n,a}^T\right],\qquad\pi_{n,a}^t \in \left\lbrace  0,1 \right\rbrace \text{.}
\end{equation}

This binary vector specifies the time intervals during which the member agrees for its appliance to be scheduled (1:~permitted, 0: not permitted). If the total predetermined energy consumption of appliance $a$ is $E_{n,a}$, the resulting power profile constraints are:
\begin{equation}
\boldsymbol{\pi}_{n,a} \cdot \mathbf{x}_{n,a}^{\mathbf{T}} \cdot \Delta t = E_{n,a}
\end{equation}
\begin{equation}
\mathrm{NOT}(\boldsymbol{\pi}_{n,a}) \cdot \mathbf{x}_{n,a}^{\mathbf{T}} = 0\text{.}
\end{equation}

Besides, loads are subject to power consumption constraints. Some appliances have predetermined consumption cycles (e.g. washing machine) while others have modular cycles (e.g. electric vehicles). For the sake of simplicity and without loss of generality, it is considered that flexible consumption is fully modular. Hence, each appliance is limited only by a maximum power level $M_{n,a}$:
\begin{equation}
0 \preceq \mathbf{x}_{n,a} \preceq M_{n,a}\text{.}
\end{equation}

In our simplified model, ESS are characterized by a maximum energy capacity $K_n$ as well as maximum power levels in charge $M_{n}^{\mathrm{ch}}$ and discharge $M_{n}^{\mathrm{dis}}$ modes:
\begin{IEEEeqnarray}{c}
0\preceq \mathbf{e}_{n,\mathrm{ind}} + \mathbf{e}_{n,\mathrm{host}} \preceq K_n  \\
-M_{n}^{\mathrm{dis}} \preceq \mathbf{s}_{n,\mathrm{ind}} + \mathbf{s}_{n,\mathrm{host}} \preceq M_{n}^{\mathrm{ch}}\text{,}
\end{IEEEeqnarray}
where $\mathbf{e}_{n,\mathrm{ind}} \triangleq \left[ e_{n,\mathrm{ind}}^1 ,\ldots, e_{n,\mathrm{ind}}^T \right]$ is the available stored energy for personal use and $\mathbf{e}_{n,\mathrm{host}} \triangleq \left[ e_{n,\mathrm{host}}^1 ,\ldots, e_{n,\mathrm{host}}^T \right]$ the available stored energy on behalf of other members, whose elements are respectively obtained by:
\begin{IEEEeqnarray}{c}
e_{n,\mathrm{ind}}^t = e_{n,\mathrm{ind}}^{t-1} + \Delta t \cdot s_{n,\mathrm{ind}}^t; \quad e_{n,\mathrm{ind}}^t \geq0 \quad \forall t \in \mathcal{T} \\
e_{n,\mathrm{host}}^t = e_{n,\mathrm{host}}^{t-1} + \Delta t \cdot s_{n,\mathrm{host}}^t; \quad e_{n,\mathrm{host}}^t \geq0 \quad \forall t \in \mathcal{T}\text{.} \quad
\end{IEEEeqnarray}
The sum of the two vectors, $\mathbf{e}_{n,\mathrm{ind}}+\mathbf{e}_{n,\mathrm{host}}=\mathbf{e}_{n,\mathrm{tot}}$, hence defines the actual ESS state of charge. Additionally, the sum of charging power vectors attributed from the pool of excess storage space to each member, $s_{n,\mathrm{mut}}$, must equal the sum of the hosting charges in the batteries $s_{n,\mathrm{host}}$, that is:
\begin{equation}
    \sum_{n \in\mathcal{N}}\mathbf{s}_{n,\mathrm{mut}} = \sum_{n \in\mathcal{N}}\mathbf{s}_{n,\mathrm{host}}\text{.}
\end{equation} 
Indeed, whereas the first sum represents a pool of virtual storage power in the eyes of the members in need, the second traces the actual physical flows of the contributing members in excess.
Note that the attributed shared ESS energy $\mathbf{e}_{n,\mathrm{mut}} \triangleq \left[ e_{n,\mathrm{mut}}^1 ,\ldots, e_{n,\mathrm{mut}}^T \right]$ is obtained by:
\begin{equation}\label{eq_e_mut2}
e_{n,\mathrm{mut}}^t = e_{n,\mathrm{mut}}^{t-1} + \Delta t \cdot s_{n,\mathrm{mut}}^t; \quad e_{n,\mathrm{mut}}^t \geq 0 \quad \forall t \in \mathcal{T}\text{.}
\end{equation}

\begin{figure}[!t]
\centering
\includegraphics[scale=0.36]{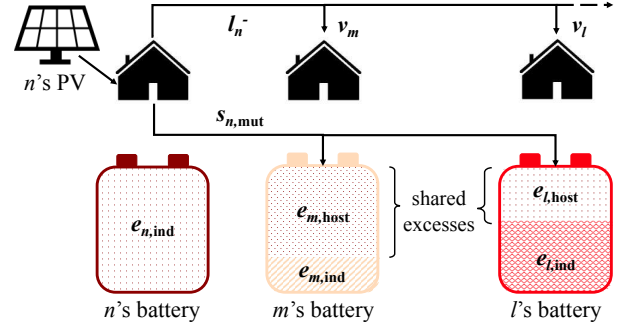}
\caption{Summary example of the variables involved in excess resources mutualization.}
\label{fig:battery}
\end{figure}

Similarly, constraint \eqref{eq_v_mut} limits the available shared power to the total outflows of the members (negative net flows):
\begin{equation}\label{eq_v_mut}
    \sum_{n \in \mathcal{N}} v_n^t \preceq \sum_{n \in \mathcal{N}} l_n^{t-}, \qquad \forall t \in \mathcal{T}\text{,}
\end{equation}
where $l_n^{t-} = \mathrm{max}(-l_n^t,0)$ is the negative part of $l_n^t$. 

Finally, the excess of generated power is available only for the members who have a positive net load, i.e.:
\begin{equation}
v_n^t \cdot l_n^{t-} = 0, \qquad \forall t \in \mathcal{T}\text{.}
\end{equation}

\textit{Summary example to highlight the practical interest of considering virtual energy exchanges}:
Let's consider a LV feeder connecting different end-users, among which member $n$ owns both a battery and a PV system (c.f. Fig. \ref{fig:battery}). It is assumed that its production exceeds its consumption. At a given time step (typically, after mid-afternoon), it may happen that the local production exceeds the consumption, while the storage space $e_{n, \mathrm{ind}}$ is full (from the energy absorbed in previous hours). In traditional market designs, the energy surplus ($s_{n,\mathrm{mut}} \cdot \Delta t$) is either consumed by local loads or transmitted outside the community. However, in our proposed framework, the energy can be stored in the batteries of other members ($m$ and $l$ in this example). In turn, these consider that stored energy respectively as $e_{m,\mathrm{host}}$ and $e_{l,\mathrm{host}}$. Although they coincide to physical flows for members $m$ and $l$, they are transparent for their billing as only the virtual flows intervene as detailed in Section \ref{sub:cost_structure}. The rest of the surplus corresponds to a negative virtual net load $l_n^-$ that benefits partly to the other members ($v_m$ and $v_l$) if they have a positive virtual net load. Our framework therefore improves the value of local generation and flexibility at the community level. 

\subsection{Power Network model}
In order to include the costs associated to power exchanges inside the community as well as to comply with the related physical constraints, the optimal power flow equations of the radial low voltage network are considered. In this work we adapt the DistFlow equations of \cite{Baran1989} to a Second Order Cone Programming (SOCP) formulation, which allows to relax the original non-convex power flow equations with a good compromise as it takes into account power losses while keeping a good computational tractability.
Each member $n$ of the community is associated to a node of the network. This node is connected to a set of adjacent nodes denoted by $\mathcal{Y}_n$. Let's introduce the variables $\varphi_{n,j}= \vert i_{n,j} \vert^2$, the squared branch current flowing between nodes $n$ and $j$, and $w_n = \vert v_n \vert^2$, the squared voltage magnitude at node $n$, where $i_{n,j}$ and $v_n$ are complex substitutes. The time superscripts $t$ have been omitted for the sake of clarity. Overall, the SOC equations of power flows for each time step $t$ are:
\addtocounter{equation}{1}
\begin{IEEEeqnarray}{ll}
p_n = \sum_{j\in \mathcal{Y}_n} p_{n,j}, &\forall n \in \mathcal{N}  \IEEEyessubnumber*\\
q_n = \sum_{j\in \mathcal{Y}_n} q_{n,j}, &\forall n \in \mathcal{N}\\
p_{n,j}^2 + q_{n,j}^2 \leq w_n \varphi_{n,j}, \qquad &\forall n,j\in \mathcal{Y}_n \label{eq:OPFc}\\
p_{n,j} + p_{j,n} = R_{n,j} \varphi_{n,j}, \qquad &\forall n,j\in \mathcal{Y}_n \\
q_{n,j} + q_{j,n} = X_{n,j} \varphi_{n,j}, \qquad &\forall n,j\in \mathcal{Y}_n \\
w_j = w_n - 2\left(R_{n,j} p_{n,j} + X_{n,j} q_{n,j} \right) \nonumber \\
\qquad + \left(R_{n,j}^2 + X_{n,j}^2\right) \varphi_{n,j}, \qquad\qquad\qquad &\forall n,j\in \mathcal{Y}_n\text{,} \qquad
\end{IEEEeqnarray}
where $p_n$ and $q_n$ are the active and reactive power injected at node $n$, $p_{n,j}$ and $q_{n,j}$ are the active and reactive branch power flowing from $n$ to $j$, while $R_{n,j}$ and $X_{n,j}$ are respectively the resistance and reactance of the branch connecting $n$ and $j$.

The equality constraint from the original Optimal Power Flow (OPF) is relaxed into an inequality constraint (\ref{eq:OPFc}), hence including the cone interior space and ensuring convexity. Both terms of the equation should tend to be equal as the objective function tends to minimize power losses (Section \ref{sec:PES}).
The link between the prosumer model and the electrical network is described by the following equation:
\begin{equation}
p_n^t = d_n^t + \sum_{a\in\mathcal{A}_{n}} x_{n,a}^t + s_{n,\mathrm{ind}}^t + s_{n,\mathrm{host}}^t\text{.}
\label{eq:physical}
\end{equation}

In contrast with (\ref{eq:netcons}), the variables $v_n^t$ and $s_{n,\mathrm{mut}}^t$ coming from the mutualization of energy surpluses are not included as they are not physically linked to the prosumer $n$. In turn, if the prosumer $n$ has available storage space, the energy from mutualized excess storage can potentially be hosted on its device, which is captured by the variable $s_{n,\mathrm{host}}^t$. The load $p_n^t$ in (\ref{eq:physical}) is referred as the physical net load on Fig. \ref{fig:community_bill}.

\subsection{Cost structure}
\label{sub:cost_structure}
The day-ahead power exchange scheduling problem is based on the minimization of the electricity costs, which consist in three contributions in the objective function:\\
\begin{enumerate}[label=\roman*)]
\item \textit{The commodity costs:} they are associated to the marginal costs of generation, and are billed by the suppliers. Each member of the community can freely choose its supplier $i\in \mathcal{S} \triangleq \{1,\ldots,n_s\}$. However, based on the consumption history and the prosumer profile, it is easy to recommend or automatically attribute the best suited supplier. It is assumed that energy is billed by suppliers according to a dynamic pricing scheme \cite{Hupez2018}. Prices are not considered to be capacity-based, i.e. they are independent of the load level, as the suppliers have a certain degree of control on their costs via their generation portfolio and market positions. This has also the advantage to provide a more legible comparison between suppliers. Besides, only the positive part of the individual net load $l_n^{t+} = \mathrm{max}(l_n^t,0)$ is billed (the negative part $l_n^{t-}$ is valorized among the community members). 
Therefore, supplier $i$ applies for each of its customer $n\in \mathcal{N}_i$:
\begin{equation}
C_{\mathrm{supp},i}^t(l_n^t) =   \gamma_{\mathrm{supp},i}^t \cdot \Delta t \cdot l_n^{t+}\text{,}
\end{equation} 
where $\gamma_{\mathrm{supp},i}^{t}$ is the commodity price (in \euro{}/kWh) of supplier $i$ at time $t$.
Within a cooperative community, the commodity costs associated to each individual are influenced by the internal transactions between members. To fairly monetize this mutualization of resources, it may be necessary that the final contribution of client $n$ differs from $C_{\mathrm{supp},i}^t(l_n^t)$. In that respect, two different ways to distribute the total costs $\sum_{n \in \mathcal{N}} C_{\mathrm{supp},i}^t(l_n^t)$ at the community level are presented in Section \ref{sub:cost_distribution}. 
It should be noted that this costs distribution (within the community) does not affect the final revenues of the supplier $i$, i.e. $ \sum_{n\in \mathcal{N}_i} C_{\mathrm{supp},i}^t(l_n^t)$. \\
\item \textit{The upstream grid costs:} they are associated to the power flows at the interface of the community, thus reflecting the effects on transmission and upstream distribution grids. These costs $C_{\mathrm{up}}^{t} \left( p_0^t \right)$ are not proportional to the load. Hence, we consider a non-linear cost function depending on the power at the interface node $p_0^t$. Without loss of generality, we consider a quadratic cost function: 
\begin{equation}
C_{\mathrm{up}}^t(p_0^t) =  \gamma_{\mathrm{up}} \cdot \left( \Delta t \cdot p_0^{t} \right)^2\text{,}
\end{equation} 
where $\gamma_{\mathrm{up}}$ represents the upstream grid fees (in \euro{}/kWh$^2$). \\
\item \textit{The local grid costs:} they are directly based on the power exchanges inside the community. We assume that each member of the community is connected to the local distribution network (cf. Fig. \ref{fig:community_bill}). Two different cost items are distinguished. One is based on the power losses in the lines while the second one is based on the line power flows in order to account for the wear and tear of the assets. The cost associated to each branch, formed by a pair of nodes $(j,k) \in \mathcal{B}$, where $\mathcal{B}$ is the set of all branches of the low voltage entity, is expressed by: 
\begin{IEEEeqnarray}{cl}
C_{\mathrm{loc}}^t(p_{j,k}^t) = &\Delta t \left( \gamma_{\mathrm{loss}} \left( p_{j,k}^t + p_{k,j}^t \right) \right.\nonumber \\
& \left. + \gamma_{\mathrm{flow}} \cdot \delta_{j,k} \cdot \vert p_{j,k}^t \vert \right)\text{,} \quad 
\end{IEEEeqnarray}
where $\gamma_{\mathrm{loss}}$ and $\gamma_{\mathrm{flow}}$ are respectively the costs of losses and line power flows (in \euro{}/kWh), while $\delta_{j,k}$ is the distance between the two nodes $j$ and $k$.\\
\end{enumerate} 

Overall, cooperating at the community level allows to positively influence all three costs contributions, i.e. through a joint consideration of (i) nodal energy exchanges (for commodity costs), and (ii)-(iii) line power flows (for grid-related fees).

\section{Power exchange scheduling}
\label{sec:PES}
\subsection{Optimization problem}
The day-ahead optimization problem minimizes the total cost of electricity consumption by the community. The benefits over a passive consumption or an individual minimization are the following:
\begin{itemize}
\item By jointly planning their consumption, end-users can find the cost-optimal trade-off between arbitrage in the dynamic pricing (i.e. shift the load when prices are low), the upstream grid costs (which are minimized by smoothing the total load over time) and the local grid costs.
\item By sharing their excess storage, they provide more flexibility to other community members, hence giving more potential to the first consideration above. Moreover, sharing excess energy from non-dispatchable sources not only provides free energy to other members but also decreases the grid costs by reducing the imports.
\item By properly coordinating storage systems and power exchanges, power flows inside the community are contained, hence decreasing the related costs.
\end{itemize}

In practice, these three benefits are deeply correlated. The objective function consists thus in jointly minimizing the following total community costs:
\begin{IEEEeqnarray}{l}
 f(\Theta)= \sum_{t=1}^T \Delta t \left[ \sum_{i \in \mathcal{S}} \left( \gamma_{\mathrm{supp},i}^t \sum_{n\in \mathcal{N}_i} l_n^{t+} \right) + \gamma_{\mathrm{up}} \cdot \Delta t \cdot \left( p_0^t \right)^2 \right. \nonumber \\
\left. + \sum_{(j,k)\in \mathcal{B}}\left( \gamma_{\mathrm{loss}} \left( p_{j,k}^t + p_{k,j}^t \right) + \gamma_{\mathrm{flow}} \left(p_{j,k}^{t+} + p_{j,k}^{t-} \right)  \delta_{j,k} \right) \right]\text{,}\quad
\label{eq:obj}
\end{IEEEeqnarray}
where $\Theta = \cup_{n\in\mathcal{N}}\Theta_n$ is the set of all decision variables including: $x_{n,a}$, $s_{n,\mathrm{ind}}$, $s_{n,\mathrm{mut}}$, $s_{n,\mathrm{host}}$ and $v_n$ for each member 
$n$. The absolute active power value in (\ref{eq:obj}) is expressed using $p_{j,k}^+ = \mathrm{max}(p_{j,k},0)$ and $p_{j,k}^- = \mathrm{max}(-p_{j,k},0)$.

The day-ahead Power Exchange Scheduling (PES) problem is therefore solved by:
\begin{IEEEeqnarray}{l}
\min_{\Theta} f(\Theta) \text{ as in (\ref{eq:obj})}  \nonumber \\
\text{s.t. (7), (9)-(20)} \text{,}
\label{eq:optim}
\end{IEEEeqnarray}
which is a convex optimization problem as both the constraints and the objective function are convex. It can be solved by standard algorithms such as interior-point or subgradient methods \cite{boyd_vandenberghe_2004}. However, it can also be conveniently solved using a distributed algorithm, thereby avoiding to resort on a third-party to guarantee the safe operation of the proposed design.

\subsection{Cost distribution}
\label{sub:cost_distribution}
Computing the minimum total cost of the community may not provide sufficient leverage for committing prosumers. Their collaboration should be incentivized by an adequate cost distribution (ensuring that all individuals have a financial interest in helping the community). In this work, we investigate two different options to distribute the cost savings arising from the global optimization \eqref{eq:optim} among community members. It should be noted that introducing network fees in the definition of the cost distribution is inherently unfair as the contribution of individuals strongly depends on their geographical location and the grid infrastructure. Instead, we choose to distribute costs based on the commodity item, accounting for the consumption and sharing behaviors of the individuals. The network costs are then passed on proportionally to their net load in the second distribution scheme.

\subsubsection{Nash equilibrium}
The first possibility consists in defining a cost distribution leading to a natural consensus. To that end, we exploit the fact that is possible to define a Nash equilibrium problem while keeping the global optimum solution \cite{Hupez2018}. It has the advantage of not requiring a central authority nor any supervision. 
Hence, based on \cite{Hupez2018}, the total cost is distributed according to a proportional key accounting for the optimum commodity costs they would selfishly obtain without cooperation:
\begin{equation}
b_n = \dfrac{C_{\mathrm{supp},i}^{\{n\}*}}{\sum_{m\in \mathcal{N}}C_{\mathrm{supp},i}^{\{m\}*}} C_{\mathrm{tot}}^{\{\mathcal{N}\}*}\text{,}
\label{eq:billing}
\end{equation}
where:
\begin{itemize}
    \item $C_{\mathrm{supp},i}^{\{n\}*}$ is the commodity cost of prosumer $n$, customer of supplier $i$, when it optimizes its load selfishly (without cooperation).
    \item $C_{\mathrm{tot}}^{\{\mathcal{N}\}*}$ is the solution of the optimization problem \eqref{eq:optim}.
\end{itemize}
Note that the superscripts specify the coalition involved and if the optimal value is considered ($^*$) or not. Besides, all the costs considered for billing derive from the computations across all time steps. 

The billing problem \eqref{eq:billing} can also be expressed as the solution of the following Nash equilibrium problem:
\begin{itemize}
\item Players: all $N$ community members
\item Strategies: all the possible strategies of the decision variable set $\Theta_n$
\item Payoffs: 
\begin{equation}
P(\Theta_n;\Theta_{-n}) = -b_n = -\dfrac{C_{\mathrm{supp},i}^{\{n\}*}}{\sum_{m\in \mathcal{N}}C_{\mathrm{supp},i}^{\{m\}*}} f(\Theta)\text{,}
\label{eq:payoff}
\end{equation}
with $\Theta_{-n} \triangleq [\Theta_1,\ldots,\Theta_{n-1},\Theta_{n+1},\ldots,\Theta_N] = \Theta\setminus\{\Theta_n\}$, the vector containing all the decision variables except those of $n$.
\end{itemize}

Given the strict convexity of (\ref{eq:payoff}), the Nash equilibrium exists and is unique \cite{Hupez2018}. Consequently, the solution can be obtained by solving (\ref{eq:billing}), which requires the computation of $N+1$ problems: $N$ selfish problems $C_{\mathrm{supp},i}^{\{n\}*}$ and the main cooperative optimization $C_{\mathrm{tot}}^{\{\mathcal{N}\}*}$.
Overall, the Nash equilibrium cannot reward participants according to their contribution in the savings. The solution tends to be more egalitarian (i.e. the costs are equally distributed among clients) and could thus be considered less incentivizing (since the offered flexibility is not properly valued at the individual level, but shared among all members). To alleviate this effect, we also study the cost distribution resulting from a derivative of the Shapley value.

\subsubsection{Shapley value}
The Shapley value is a concept of cooperative game theory assigning a share of the gains obtained by forming coalition of different actors \cite{Shapley1953}. 
It allows to remunerate end-users with payoffs that correspond to their individual contributions to the gains of the community, and is characterized by a collection of desirable properties introducing a certain degree of fairness. 
Let $\mathcal{Q} \subseteq \mathcal{N} $, define a coalition of size $\vert \mathcal{Q} \vert$ and the characteristic function $ v( \mathcal{S} ):2^{\mathcal{N}}\rightarrow \mathbb{R} $ describe the savings associated to the formation of $\mathcal{Q}$. A cooperation game is defined by the pair $(\mathcal{N}, v)$ and the Shapley value $\phi_n$ attributed to a given player $n$ is given by
\begin{equation}
\phi_n(v) = \sum_{\mathcal{Q} \subseteq \mathcal{N}\setminus \{n\}} \dfrac{\vert \mathcal{Q} \vert! (N - \vert \mathcal{Q} \vert - 1)!}{N!} [v(\mathcal{Q}\cup\{n\}) -v(\mathcal{Q})] \text{.}
\end{equation}

Although all the $2^N$ combinations of coalition are necessary to compute the Shapley values, it should be noted that the only coalition possible in the cost structure considered in this paper is the grand coalition, that is $\mathcal{N}$. Indeed, as previously mentioned, the network costs cannot be directly echoed to an individual or a subset of the community as the power flows depend on the joint load of all clients. Based on all the previously mentioned considerations, we propose a billing with a double distribution key. On one hand, the commodity costs are distributed according to their Shapley value (numerator of the first term in \eqref{eq:bill}) obtained on the optimum commodity cost and scaled to the actual commodity costs (remaining of the first term). On the other hand, the grid costs are allocated proportionally to the net load. The total costs $b_n$ incurred by participant $n$ are thus:
\begin{equation}
b_n = \dfrac{C_{\mathrm{supp},i}^{\{n\}*}- \phi_n(v)} {C_{\mathrm{supp}}^{\{\mathcal{N}\}*}} C_{\mathrm{supp}}^{\{\mathcal{N}\}}  + \dfrac{l_n}{\sum_{m\in \mathcal{N}}l_m} C_{\mathrm{grid}}^{\{\mathcal{N}\}}\text{,}
\label{eq:bill}
\end{equation}
where :
\begin{itemize}
\item  $C_{\mathrm{grid}}^{\{\mathcal{N}\}}$ is the sum of $C_{\mathrm{up}}^{\{\mathcal{N}\}}$ and $C_{\mathrm{loc}}^{\{\mathcal{N}\}}$, the total upstream and local grid costs.
\item $C_{\mathrm{supp}}^{\{\mathcal{N}\}*}$ is the optimum community commodity cost obtained by solving a truncated version of \eqref{eq:optim} which considers commodity costs only (i.e. excluding network costs).
\item $C_{\mathrm{supp}}^{\{\mathcal{N}\}}$ and $C_{\mathrm{grid}}^{\{\mathcal{N}\}}$ are respectively the actual commodity and grid costs of the community. They are obtained by extracting respectively the commodity and grid costs after solving the full version of \eqref{eq:optim}. 
\end{itemize}
and with $C_{\mathrm{supp}}^{\{\mathcal{N}\}} + C_{\mathrm{grid}}^{\{\mathcal{N}\}}=C_{\mathrm{tot}}^{\{\mathcal{N}\}*}$, the solution of the optimization problem \eqref{eq:optim}. Note that the computation of the $2^N$ combinations of coalition costs is sufficient to calculate the bills. 

\subsubsection{Benchmark}\label{section:bench}
In order to quantify the added value of both cost distributions, i.e. (i) the Nash equilibrium, and (ii) the Shapley-based payoffs, we introduce a proper benchmark that quantifies what can be achieved using individualized optimal policies (without collaboration). To that end, we consider a framework (iii) where each end-user individually minimizes its cost objective function. Practically, two different individual strategies are analyzed (in Section \ref{sec:case_studies}), which respectively account for the minimization of the commodity costs and the minimization of the grid costs over the daily scheduling horizon. Table \ref{tab:distribution} summarizes the distribution keys of the three methodologies. 
\begin{table}[h!]
\centering
\caption{Distribution key summary.}
\begin{tabular}{|l|ccc|}
\hline
           & Nash (i) & Shapley (ii) & Indiv. strat. (iii) \\ \hline
$C_{\mathrm{supp}}^{\{\mathcal{N}\}}$ &\multirow{2}{*}{$\dfrac{C_{\mathrm{supp},i}^{\{n\}*}}{\sum_{m\in \mathcal{N}}C_{\mathrm{supp},i}^{\{m\}*}}$} &$\dfrac{C_{\mathrm{supp},i}^{\{n\}*}- \phi_n(v)}{C_{\mathrm{supp}}^{\{\mathcal{N}\}*}}$ & $\dfrac{C_{\mathrm{supp},i}^{\{n\}*}}{\sum_{m\in \mathcal{N}}C_{\mathrm{supp},i}^{\{m\}*}}$          \\
$C_{\mathrm{grid}}^{\{\mathcal{N}\}}$ & & $\dfrac{l_n}{\sum_{n\in \mathcal{N}}l_n}$      & $\dfrac{l_n}{\sum_{n\in \mathcal{N}}l_n}$  \\\hline
\end{tabular}
\label{tab:distribution}
\end{table}

Regarding both cooperation schemes (i)-(ii), it is necessary that participants pledge their contribution and be transparent about their engagement for the predefined consensus. This consideration could be treated by the use of a blockchain.

\section{Case studies}
\label{sec:case_studies}

\begin{figure}[!b]
\centering
\includegraphics[scale=1]{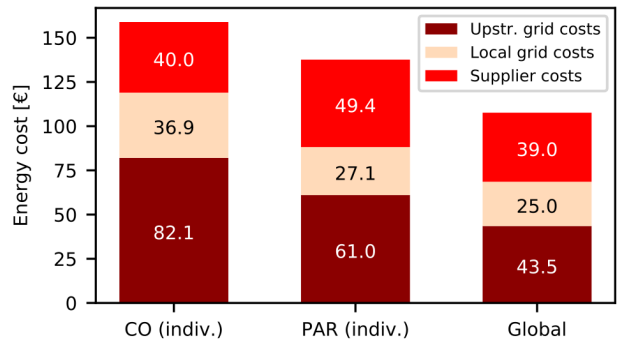}
\caption{Comparison of the total cost division for different load management scenarios.}
\label{fig:cost_approaches}
\end{figure}
\begin{figure*}[!t]
\centering
\includegraphics[scale=1]{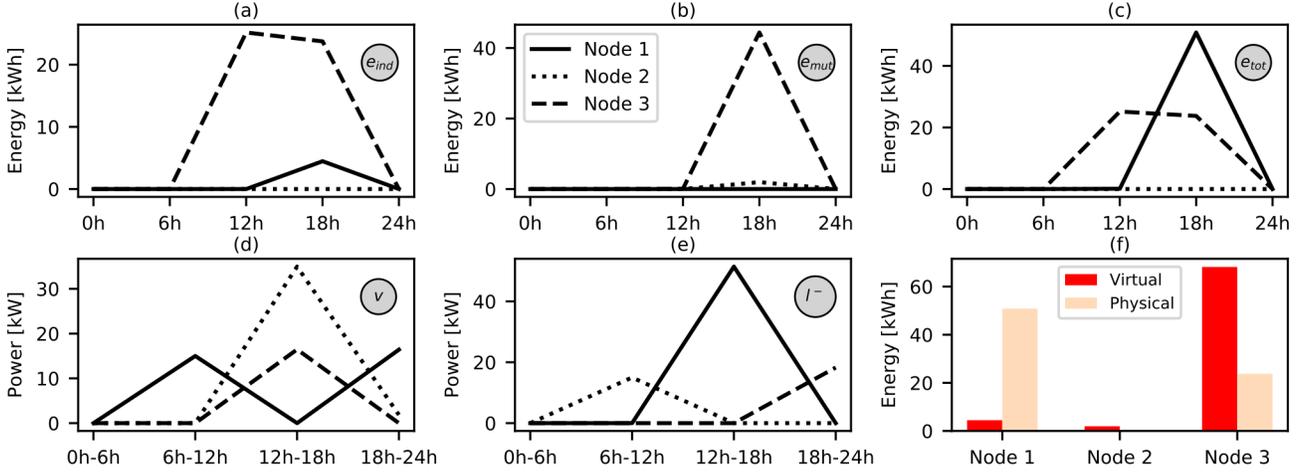}
\caption{Temporal evolution of the personal available energy stored locally $\mathbf{e_{n,\mathrm{ind}}}$ (a), the personal available energy stored remotely $\mathbf{e_{n,\mathrm{mut}}}$ (b), the total energy stored locally $\mathbf{e_{n,\mathrm{tot}}}$ (c), the allocated portion of excess power $\mathbf{v_{n}}$ (d) and the virtual negative net load $\mathbf{l_{n}^-}$ (e) $\forall n$. Virtual (available for personal use) and physical ESS state of charge at 18h (f).}
\label{fig:scheduling}
\end{figure*}

The proposed framework is analyzed on a benchmark composed of $N$ = 3 prosumer nodes. Each of them is assigned to a load representing the equivalent of about five low voltage users, hence summing up to 15 users, so as to reflect the conditions that could be found on a residential feeder (typical Belgian feeder such as in \cite{Klonari2015}) and obtain representative power flows in the LV network. The feeder is assumed to be one kilometer long and is fully radial. 
Table \ref{tab:load} summarizes the main features of the nodal loads, i.e. details about their electrical equipment, their choice of supplier (1: cheaper nights, 2: cheaper days) and their potential in terms of load flexibility. Note that in order to ease the observations, the scheduling takes place for a day comprising four time slots. Additional information regarding the LV network, the load constraints and the tariffs applied can be found in the online appendix \cite{Appendix}.
\begin{table}[h]
\centering
\caption{Summary of the end-users characteristics.}
\begin{tabular}{|l|cccc|}
\hline
       & \multicolumn{1}{l}{PV} & \multicolumn{1}{l}{ESS} & \multicolumn{1}{l}{Supplier} & \multicolumn{1}{l|}{Shiftability}  \\ \hline
Node 1 & X                      & X                       & 1                            & 28\%                                                             \\
Node 2 & X                      &                         & 1                            & 42\%                                \\
Node 3 &                        & X                       & 2                            & 56\%                               \\ \hline
\end{tabular}
\label{tab:load}
\end{table}

In what follows, we will firstly compare, in Section \ref{sec:4a}, the total costs (aggregated at the community level) between individual and cooperative approaches. Secondly, in Section~\ref{sec:4b}, we will analyze in more details how the proposed cooperative market design allows to cost-efficiently share the available flexibility among end-users, thus highlighting the interest of differentiating physical from virtual energy exchanges. Then, based on the scheduling solution obtained by the global optimization, we will investigate in Section~\ref{sec:4c} how the costs are distributed among individuals (using both the Nash equilibrium and the Shapley value). Finally, we will highlight in Section \ref{sub:comparison} the benefits of cooperation by comparing the global cooperative optimization with a traditional framework where each end-user optimizes its own electricity bill with fixed grid fees.

\subsection{Day-ahead scheduling of end-users}\label{sec:4a}
A comparison of the total costs for three different load management approaches are displayed in Fig. \ref{fig:cost_approaches}. 
The first two are the individual strategies briefly presented in Section \ref{section:bench} as a benchmark to quantify the added value of the global optimization. Firstly, we consider a basic situation, referred to as Commodity-Only (CO), where each individual optimizes its own commodity costs (by using its own flexibility to shift the load during off-peak prices), thus obtaining $C_{\mathrm{supp},i}^{\{n\}*} \forall n$. The second approach applies the Peak-to-Average Ratio (PAR), where end-users try to integrate network costs in their own objective function. However, since these two policies do not involve cooperation at the community level, there is a lack of global vision that limits their potential.
This effect is quantified by solving the global optimization \eqref{eq:optim}, which yields the total commodity costs of the community $C_{\mathrm{supp}}^{\{\mathcal{N}\}}$.

Interestingly, introducing the Peak-to-Average Ratio (PAR) allows to decrease the total costs in comparison with the individual minimization of commodity costs (Fig. \ref{fig:cost_approaches}). Indeed, by exclusively focusing on commodity costs, end-users tend to concentrate their energy consumption at specific time slots (when dynamic prices are the lowest). This behavior results in consumption peaks that significantly increase the grid costs. This effect is further exacerbated when all clients rely on the same incentives (i.e. when they have similar commodity tariff schemes). The PAR optimization alleviates this issue through a better homogenization of the energy exchanges along the day, hence leading to higher commodity costs but to a significant reduction of both grid fees components (within and outside the community).
Overall, it is important to emphasize that these two individual approaches probably represent optimistic individual behaviors as they assume that each individual will optimize its flexible resources on a daily basis. A more passive behavior of end-users or a lack of control means would surely yield higher costs.
These optimistic outcomes from both individual optimization are nonetheless outperformed by the global optimization \eqref{eq:optim}, since the latter takes advantage of the flexibility of all available resources. The cooperative framework achieves the minimal costs for the three contributions of the final electricity bill. This arises from the inclusion of energy sharing mechanisms that enable to optimally share inter-participant flexibility, i.e. excess mutualized storage space and photovoltaic generation. As the billing of each individual is based on the total cost, decreasing net imports of other individuals through the provided flexibility benefits the whole community.

\subsection{Benefits of the community-based optimization}
\label{sec:4b}
If we focus on the solution of the global optimization, the relevance of the different energy sharing mechanisms can be easily highlighted. For instance, node 1, whose end-users present a poor load shiftability (c.f. Table \ref{tab:load}), could not leverage its DERs (excess production and storage), available mostly in the afternoon, without sharing. Indeed, most of its load must be planned in the night and morning. In contrast, we observe that when the global optimization is considered, node 1 hosts a significant amount of energy for node 3 and a small quantity for node 2. Indeed, from (a-c) on Fig. \ref{fig:scheduling}, it can be identified that most of the energy stored at node 1 is not for personal use ($e_{1,\mathrm{tot}}^{18h} \gg e_{1,\mathrm{ind}}^{18h}$) and that $e_{2,\mathrm{mut}}^{18h}$ and $e_{3,\mathrm{mut}}^{18h}$ are both positive. Furthermore, node 2 and 3 benefit from excess energy originating from the PV generation of node 1 (cf. (d-e) on Fig. \ref{fig:scheduling}). However, beyond benefiting indirectly from the underlying decrease of the grid costs, node 1 also gets excess energy from nodes 2 and 3 respectively in the morning and evening. Note that no energy is stored at night as a significant amount of energy must be supplied and transit from the upstream grid. An additional storage load would yield a considerable increase of the associated bill.

Furthermore, from Fig. \ref{fig:scheduling}(f), the distinction between physically and virtually stored energy can be highlighted. The difference between both values correspond to either the hosted energy $e_{n,\mathrm{host}}^t$ by an individual if there is excess of physical storage space or the remote energy of an individual $e_{n,\mathrm{mut}}^t$ otherwise.

\subsection{Effects of different cost distributions}\label{sec:4c}
\begin{figure}[!h]
\centering
\includegraphics[scale=1]{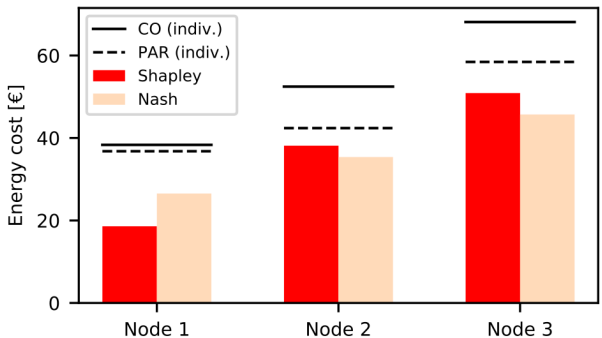}
\caption{Comparison of the final electricity bills of end-users with the different cost distribution keys.}
\label{fig:cost_distribution}
\end{figure}

After the minimization of the total costs \eqref{eq:optim}, it is necessary to fairly and optimally distribute these costs to all members of the community. Both cost distribution keys (presented in Section \ref{sub:cost_distribution}) are compared, and outcomes are illustrated in Fig. \ref{fig:cost_distribution} for the final electricity bill of the $N$ = 3 nodes of the network. For the sake of comparison, results from the CO and PAR optimizations are also depicted.

We observe that the gains arising from the PAR optimization leads to costs reduction for all three nodes of the community with respect to the sole minimization of commodity costs. All consumers have thus a financial incentive for incorporating grid fees into their day-ahead scheduling. However, these gains are much lower for node 1, which can be explained by the fact that both CO and PAR optimizations result in a similar planning solution (they schedule most of the flexible load when the PV injection is high). 
Overall, although improving the way end-users allocate their own resources can be profitable, these are incentivized into forming local communities to further reduce their electricity bill. 

When global optimization is considered, the two cost distributions yield different trends. Under the present scenario, the resulting gains under the Nash equilibrium solution \eqref{eq:billing} generate smaller costs differences between community members than the outcome obtained with the Shapley value. Indeed, the latter option tends to reward actors who contribute most to the decrease of the commodity whereas the Nash equilibrium solution values the potential flexibility. In that regard, the significant contribution of node 1 highlighted in Section \ref{sec:4b} is better valued with the Shapley-based payoffs. The two options thereby give different incentives to fully leverage the flexibility of LV prosumers.  
\begin{table}[h]
\centering
\caption{Impact of storage on costs}
\label{tab:storage}
\begin{tabular}{|c|cccc|}
\hline
Energy cost {[\euro{}]} & Node 1 & Node 2 & Node 3 & Total                   \\ \hline
ESS / Nash            & 24.25  & 35.97  & 47.42  & \multirow{2}{*}{107.64} \\
ESS / Shapley         & 19.75  & 36.39  & 51.50  &                         \\ \hline
No ESS / Nash              & 26.7   & 33.70  & 49.07  & \multirow{2}{*}{109.47} \\
No ESS / Shapley           & 21.38  & 36.49  & 51.60  &                         \\ \hline
\end{tabular}
\end{table}

These observations can be further highlighted by modifying the available storage capacity. Table \ref{tab:storage} gives the impact of storage on the energy cost for each node. From this table, we observe that if there is no available storage for nodes 1 and 3, the total electricity cost of the community is increased. In particular, the solution given by the Nash equilibium raises the electricity bill of both clients (1 and 3), while the Shapley-based outcome only affects node 1. Indeed, node 1 has less load flexibility than other members, and is thereby more impacted by the Shapley methodology that rewards the actual mobilized flexibility. It can be also noted that the absolute amount of the bill with zero ESS capacity could be much greater during a day with higher PV production. This results from the increased grids costs related to the higher injection.

\subsection{Benefits of sharing the grid fees at the community level}
\label{sub:comparison}
In this section, the outcomes obtained with our proposed cooperative approach (which properly considers non-linear grid costs) is compared with a benchmark representative of the current reality in most places worldwide. Hence, the latter design assumes an individual approach where each end-user optimizes its own cost function, in which grid costs are linearly integrated (proportional to the energy costs). 

Table IV shows, for the same initial benchmark conditions, the relative differences of billings for various fixed grid rates (price per unit of consumed energy) in comparison to our cooperative approach with aggregative non-linear grid costs (a positive value corresponds to an increase of the costs). 

\begin{table}[h]
\centering
\caption{Costs difference between aggregative and linear grid fees.}
\begin{tabular}{|c|ccc|}
\hline
\% costs difference & Node 1 & Node 2 & Node 3 \\ \hline
0.208 \euro{}/kWh      & 14.3   & 6.5    & 5.1    \\
0.30 \euro{}/kWh        & 48.5   & 37.2   & 35.2   \\
0.10 \euro{}/kWh        & -25.8  & -29.5  & -30.2  \\ \hline
\end{tabular}
\label{tab:difference}
\end{table}

Firstly, we can observe that even the optimistic case of a fixed grid price corresponding to the mean price obtained under the global optimization (0.208 \euro{}/kWh) leads to an increase of the billing for the three considered nodes. This comes from the fact that the individual optimization leads to higher imports (no sharing of surpluses). In reality, as there is no optimization across grid costs, the incurred costs should be even greater (c.f. individual approaches on Fig. \ref{fig:cost_approaches}) and a higher fixed rate could be expected. In that regard, the results show that increasing the fixed grid price (0.30 \euro{}/kWh) has a greater impact on node 1. This arises from its lower flexibility potential that prevents it from adapting its schedule. On the other hand, lower grid fees, i.e. 0.10 \euro{}/kWh (which could reflect a more efficient planning and operation of the whole system), is smoothing the differences among nodes.

So far, the results of an aggregative non-linear grid costs framework have been exposed considering both the individual and community approaches (c.f. Sections \ref{sec:4a} to \ref{sec:4c}), while this section introduced a linear grid costs framework under the individual approach. It should be noted that a community approach using linear grid costs could have been considered, but it is not straightforward and implies structural changes.

Indeed, the aggregative grid costs framework formalizes the interdependence of individuals sharing a same asset: the grid infrastructure. This naturally promotes cooperation since all the individual problems are linked to form a global problem. In this way, the mutualizing mechanisms are driven by the optimization of the global problem. On the contrary, linear grid costs leave the individual problems separable. Introducing mutualizing mechanisms in such framework is not driven by a natural consensus. On one hand, the PV excesses, although they can be possibly shared if the actors are not penalized (e.g. no injection grid costs if surplus are shared), find no natural consensus on a distribution key. On the other hand, an individual finds no interest in sharing its surplus of storage space unless it gets a financial advantage. Another market model should therefore be developed in such case. This further highlights the benefits of our approach reflecting the truth costs through the proper consideration of the aggregative nature of grid fees and the zero-marginal cost of sharing energy surpluses.

\section{Conclusion}
\label{sub:Conclusion}
This work has presented a new practical framework to optimally leverage the flexible resources distributed throughout low voltage networks. The solution aims at alleviating issues from centralized and fully distributed paradigms by relying on local communities (using the existing infrastructure). The paper firstly introduces a global optimization tool aiming at achieving the cost-optimal allocation of available resources at the community level, accounting for both commodity and grid-related costs. Then, the total gains are distributed among end-users using two different techniques, namely a Nash equilibrium and the Shapley value. 

Results highlight the significant cost savings of forming a community with respect to the individual optimization of assets. Then, we observe that distributing the resulting costs using the Shapley value allows to endogenously reward the actors that contribute most to the costs reduction through their mobilized flexibility, whereas the Nash equilibrium tends to reward the potential flexibility consent. The two implementations thus offer different incentives to fully leverage the flexibility of the LV prosumers.

\end{document}